\documentclass[structabstract]{aa}
\usepackage{graphicx}
\usepackage{natbib}
\bibpunct{(}{)}{;}{a}{}{,}

\begin{document}

\title{The properties of ten O-type stars in the low-metallicity galaxies IC~1613, WLM and NGC~3109\thanks{Based on observations obtained at the European Southern Observatory under program IDs 085.D-0741, 088.D-0181 and 090.D-0212.}}
\author{F. Tramper \inst{\ref{api}} 
	\and H. Sana\inst{\ref{stsci}}
	\and A. de Koter\inst{\ref{api},\ref{leuven}}
	\and L. Kaper\inst{\ref{api}}
	\and O.~H. Ram{\'{\i}}rez-Agudelo\inst{\ref{api}}
	}
\institute{Anton Pannekoek Institute for Astronomy, University of Amsterdam, Science Park 904, PO Box 94249, 1090 GE Amsterdam, The Netherlands\label{api} 
	\and ESA / Space Telescope Science Institute, 3700 San Martin Drive, Baltimore, MD 21218, USA\label{stsci}
	\and Instituut voor Sterrenkunde, KU Leuven, Celestijnenlaan 200D, 3001 Leuven, Belgium\label{leuven}
	}

\authorrunning{Tramper et al.}
\titlerunning{The properties of low-metallicity O stars}

\abstract
{Massive stars likely played an important role in the reionization of the Universe, and the formation of the first black holes. They are potential progenitors of long-duration gamma-ray bursts, seen up to redshifts of about ten. Massive stars in low-metallicity environments in the local Universe are reminiscent of their high redshift counterparts, emphasizing the importance of the study of their properties and evolution. In a previous paper, we reported on indications that the stellar winds of low-metallicity O stars may be stronger than predicted, which would challenge the current paradigm of massive star evolution.}
{In this paper, we aim to extend our initial sample of six O stars in low-metallicity environments by four. The total sample of ten stars consists of the optically brigthest sources in IC1613, WLM, and NGC3109. We aim to derive their stellar and wind parameters, and compare these to radiation-driven wind theory and stellar evolution models.}
{We have obtained intermediate-resolution VLT/X-Shooter spectra of our sample of stars. We derive the stellar parameters by fitting synthetic {\sc fastwind \normalfont} line profiles to the VLT/X-Shooter spectra using a genetic fitting algoritm. We compare our parameters to evolutionary tracks and obtain evolutionary masses and ages. We also investigate the effective temperature versus spectral type calibration for SMC and lower metallicities. Finally, we reassess the wind momentum versus luminosity diagram.}
{The derived parameters of our target stars indicate stellar masses that reach values of up to 50 $M_{\odot}$. The wind strengths of our stars are, on average, stronger than predicted from radiation-driven wind theory and reminiscent of stars with an LMC metallicity. We discuss indications that the iron content of the host galaxies is higher than originally thought and is instead SMC-like. We find that the discrepancy with theory is lessened, but remains significant for this higher metallicity. This may imply that our current understanding of the wind properties of massive stars, both in the local universe as well as at cosmic distances, remains incomplete.

}
{}

\maketitle

\section{Introduction}\label{sec:intro}

\par It is expected that in the early, metal-poor Universe the formation of massive stars was favored. These stars may have played an important role in the reionization of the gas that was cooling as a result of the expansion of space \citep[e.g.,][]{haiman1997}, and produced the first black holes \citep[e.g.,][]{madau2001,micic2011}. The final collapse of single rapidly rotating massive stars in low-metallicity environments is a potential channel toward the  production of hypernovae and long-duration gamma-ray bursts \citep[e.g.,][]{yoon2005, woosley2006}.
\par The study of low-metallicity massive stars is thus crucial in our understanding of the early Universe. While the Magellanic Clouds provide access to massive stars in environments with metallicities down to 20\% of solar, for lower metallicities we have to look to more pristine dwarf galaxies in the Local Group. With 8-10m class telescopes, the stellar populations in these galaxies can be resolved, but obtaining spectra of individual massive stars hosted by these systems remains challenging and expensive in terms of observing time. Consequently, this has so far mostly been done at low spectral resolution \citep[at resolving power $R =  \lambda / \Delta \lambda \sim 1\,000-2\,000$; e.g., ][]{bresolin2006, bresolin2007, evans2007, castro2008}. 
\par The advent of X-Shooter \citep{vernet2011} on ESO's Very Large Telescope (VLT) has opened up the opportunity to observe massive stars in galaxies as far as the edge of the Local Group at intermediate resolution \citep[$R \sim 5\,000-11\,000$,][]{hartoog2012}. Apart from the better resolved shapes of the spectral lines, a higher spectral resolution facilitates a better nebular subtraction. This allows for a more detailed quantitative spectroscopic analysis. 
\par As the mass loss of massive stars through their stellar winds dominates their evolution, understanding the physical mechanism driving these winds is very important. The winds are thought to be driven by radiation pressure on metallic ion lines \citep[e.g.,][]{lucy1970, castor1975,kudritzki2000}. Consequently, the strength of the stellar winds is expected to scale with metallicity, with the prediction that $\dot{M} \propto Z^{0.69\pm0.10}$ \citep{vink2001}. This metallicity scaling has been verified empirically by \cite{mokiem2007}, who find $\dot{M} \propto Z^{0.78\pm0.17}$ for O stars in the Galaxy and Magellanic Clouds.

\begin{table}
\centering
\caption{Adopted properties of the host galaxies.}\label{tab:galaxies}
\begin{tabular}{l c c c c}
\hline\hline
Galaxy 	& d 			& $E(B-V)$ 	& $Z/Z_{\odot}$\tablefootmark{a}  & References\\
		& {\tiny (kpc)}	& 			\\
\hline \\[-8pt]
IC~1613 	&	720		& 0.025		& 0.16	& 1, 2, 3\\
WLM 	&	995		& 0.08		& 0.13	& 4\\
NGC~3109 &	1300		& 0.14		& 0.12	& 5, 6, 7\\
\hline
\end{tabular}
\tablefoot{
\tablefoottext{a}{Metallicity for IC~1613 and NGC~3109 are derived from B-supergiants and based on the oxygen abundance, adopting $12 + \log(\mathrm{O}/\mathrm{H})_{\odot} = 8.69$ \citep{asplund2009}. WLM metallicity is based on abundances of iron-group elements obtained from B-supergiants.}}
\tablebib{
 (1) \citet{pietrzynski2006}; (2) \citet{schlegel1998}; (3) \citet{bresolin2007}; (4) \citet{urbaneja2008}; (5) \citet{soszynski2006}; (6) \citet{davidge1993}; (7) \citet{evans2007}}
\end{table}

 \begin{figure}[!h]
   \resizebox{\hsize}{!}{\includegraphics{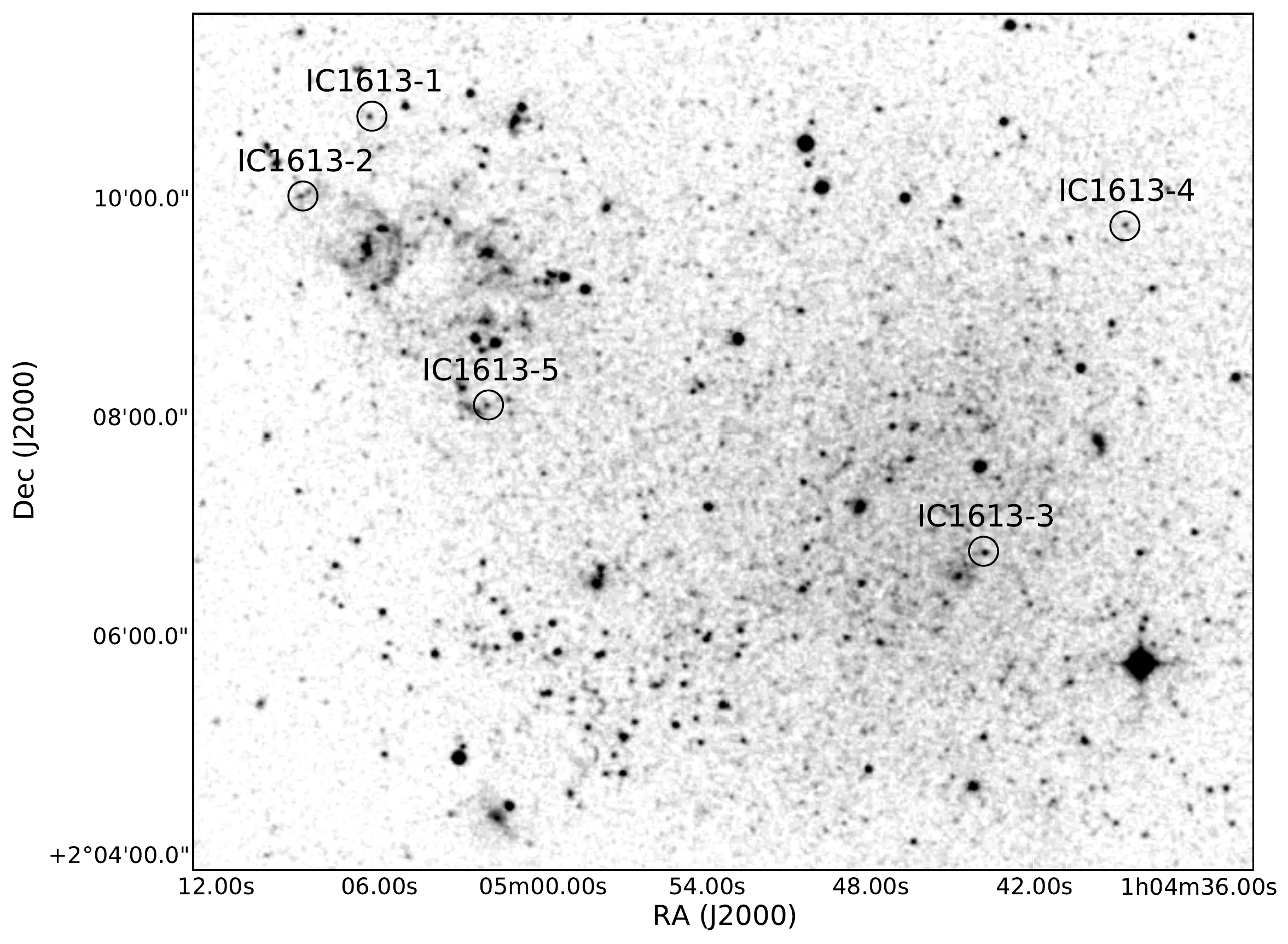}}
  \caption{Location of the target stars in IC~1613. North is up and east to the left.}
  \label{fig:ic1613}

   \resizebox{\hsize}{!}{\includegraphics{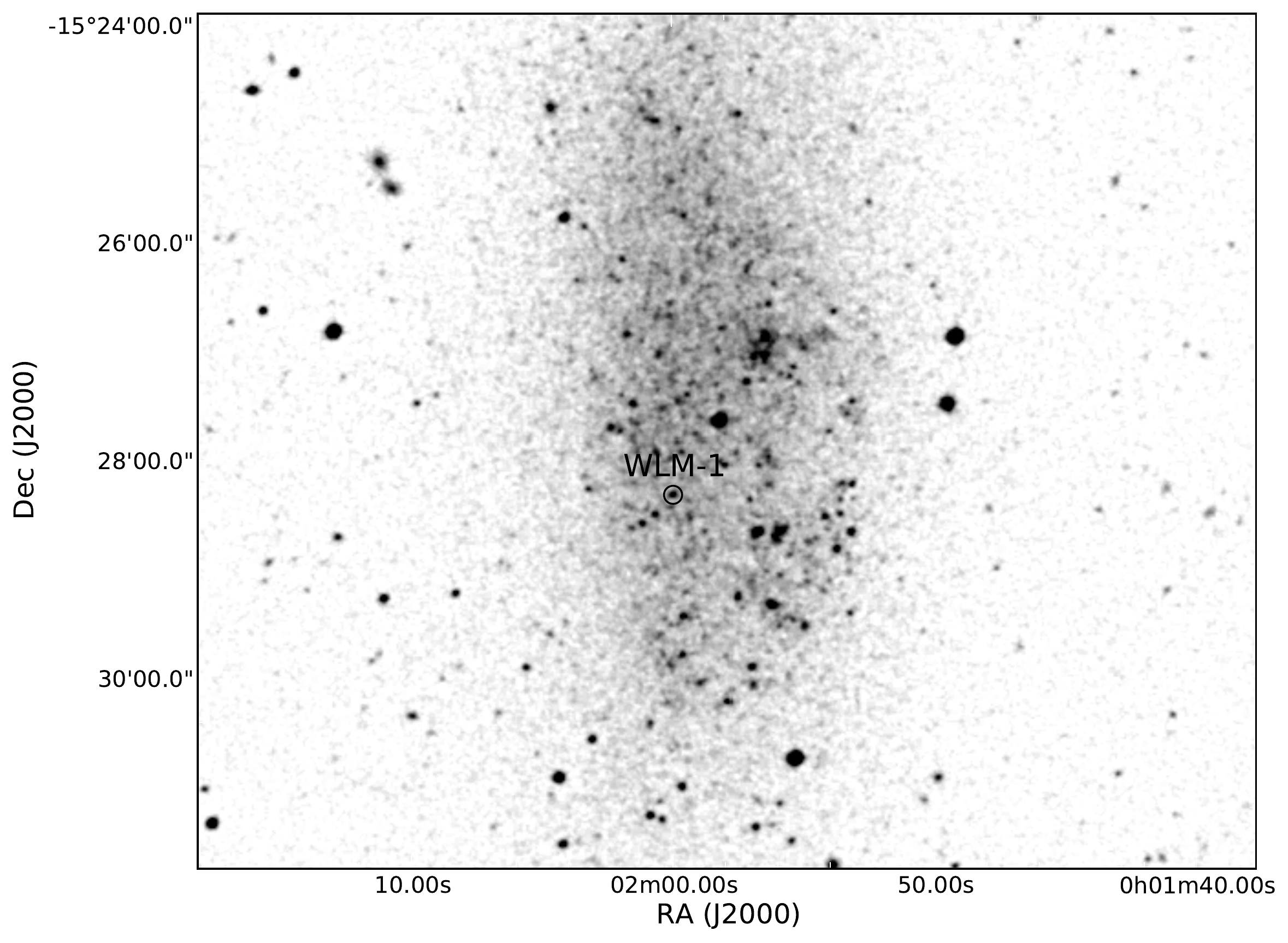}}
  \caption{Location of the target star in WLM. North is up and east to the left.}
  \label{fig:wlm}

   \resizebox{\hsize}{!}{\includegraphics{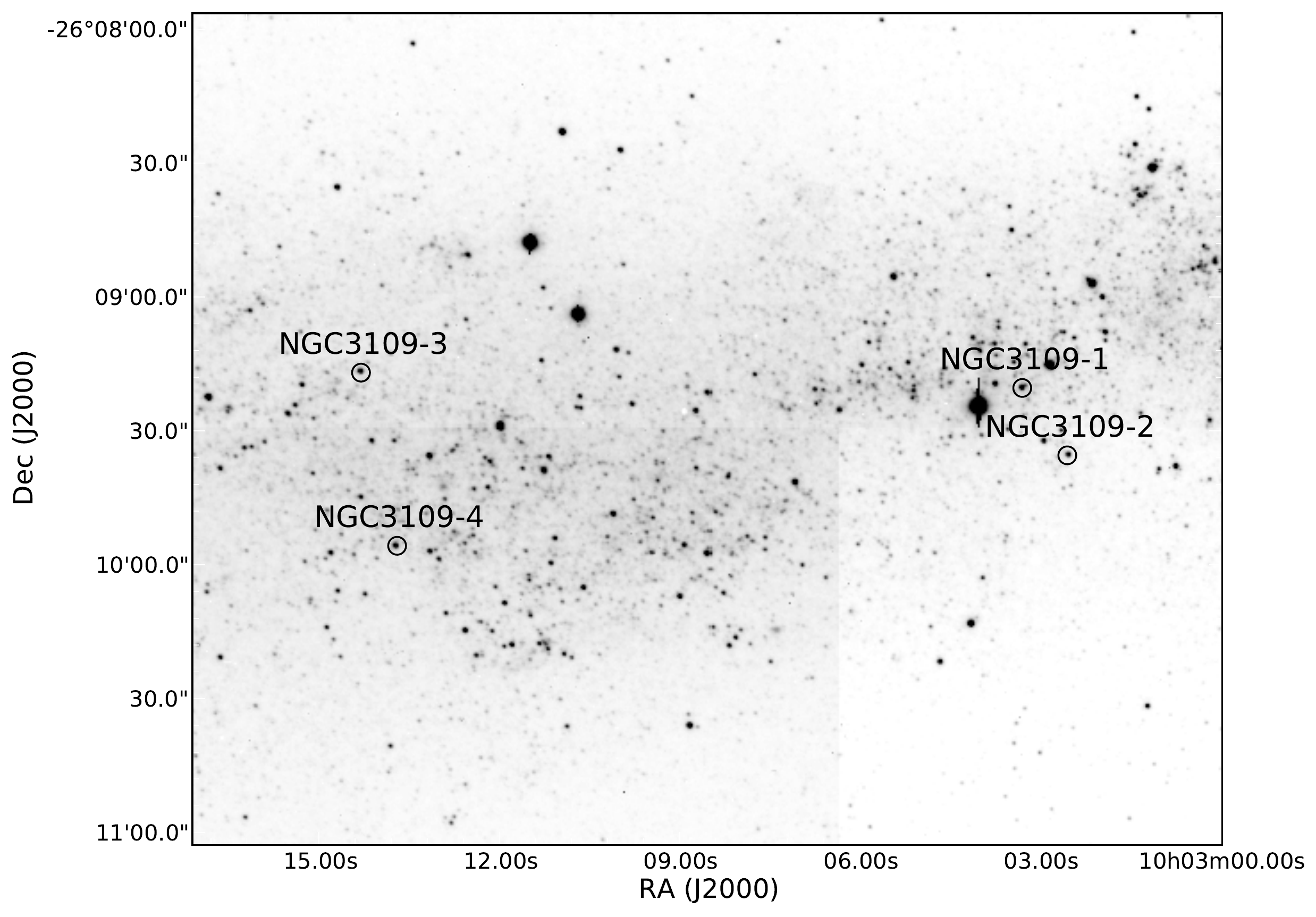}}
  \caption{Location of the target stars in NGC~3109. North is up and east to the left.}
  \label{fig:ngc3109}
\end{figure}

\par To quantify the $\dot{M}(Z)$ relation in even lower metallicity environments, we presented the first intermediate-resolution quantitative spectroscopic analysis of O-type stars with a oxygen abundance that suggests a sub-SMC metallicity in \citet[][henceforth Paper I]{tramper2011}. We unexpectedly found stellar winds that are surprisingly strong, reminiscent of an LMC metallicity. This apparent discrepancy with radiation-driven wind theory is strongest for two stars, one in WLM and one in NGC~3109. \cite{herrero2012} also report a stronger than predicted wind strength for an O-type star in IC~1613. However, observations of a larger sample of stars, as well as observations in the UV, are necessary to firmly constrain the wind properties of these stars and to prove or disprove that O stars at low metallicities have stronger winds than anticipated. 
\par A first step towards this goal has been made by \cite{garcia2014}, who obtained HST-COS spectra of several O-type stars in IC~1613, and used these to derive terminal wind velocities. They show that the wind momentum for the star analysed by \cite{herrero2012} can be reconciled with the theoretical predictions when their empirical value for the terminal velocity is adopted. They also find indications that the $\alpha$-to-iron ratio in IC~1613 may be sub-solar, which could partly explain the observed strong winds. A full analysis of the UV spectrum to constrain the mass-loss properties of the stars in their sample is still to be done.
\par In this paper, we extend our optical sample of O stars in low-metallicity galaxies by four. We constrain the physical properties of the full sample of ten O stars and reassess their winds strengths. Furthermore, we discuss the evolutionary state of the objects, that are among the visually brightest of their host galaxies. We use our results in combination with results from the literature to reassess the low-metallicity effective temperature - spectral type scale.
\par The location of all stars in our sample within their host galaxies is indicated in Figures~\ref{fig:ic1613}, \ref{fig:wlm} and \ref{fig:ngc3109}. The host galaxies are of a late type (dwarf irregulars), and have likely been forming stars continuously during their life \citep{tolstoy2009}. The distance and metallicity of the host galaxies that we adopt are given in Table~\ref{tab:galaxies} (but see Section~\ref{sec:mdot} for a discussion on the metallicities). 

\begin{table*}
\centering
\caption{Observational properties of the target stars.}\label{tab:stars}
\begin{tabular}{l l c c c c c c}
\hline\hline
 ID\tablefootmark{a} & ID & R.A. & Decl. & $V$\tablefootmark{b} & Spectral Type & $M_V$ & RV\\
 {\tiny \textit{This work}} & {\tiny \textit{Previous}} &(J2000)&  (J2000) & & & & km s$^{-1}$\\
\hline \\[-8pt]
IC1613-1 (I1)$^{1,3}$ & A13 & 01 05 06.21& +02 10 44.8 & 19.02 & O3.5 V((f)) &  $-5.55$ & $-240$\\
IC1613-2 (I2)$^{1,3}$ & A15 & 01 05 08.74 & +02 10 01.1 & 19.35 & O9.5 III & $-5.11$	& $-240$\\
IC1613-3 (I3)$^{1,3}$ & B11 & 01 04 43.82 & +02 06 46.1 & 18.68 & O9.5 I &  $-5.84$	& $-240$\\
IC1613-4 (I4)$^{1,3}$ & C9 & 01 04 38.63 & +02 09 44.4 & 19.02 & O8 III((f)) & $-5.44$	& $-265$\\
IC1613-5 (I5)$^{2,3}$ & B7 & 01 05 01.95 & +02 08 06.5 & 18.99 & O9 I &  $-5.29$	& $-214$\\
WLM-1 (W1)$^{1,4}$ & A11 & 00 01 59.97 & $-$15 28 19.2 & 18.40 & O9.7 Ia & $-6.83$\tablefootmark{c}	& $-135$\\
NGC3109-1 (N1)$^{1,5}$ & 20 & 10 03 03.22 & $-$26 09 21.4 & 19.33 & O8 I & $-6.67$	& $407$\\
NGC3109-2 (N2)$^{2,5}$ & 33 & 10 03 02.45 & $-$26 09 36.11 & 19.57& O9 If & $-6.41$	& $504$ \\
NGC3109-3 (N3)$^{2,5}$ & 34 & 10 03 14.24 & $-$26 09 16.96 & 19.61 & O8 I(f) & $-6.39$	& $415$\\
NGC3109-4 (N4)$^{2,5}$ & 35 & 10 03 13.65 & $-$26 09 55.76 & 19.70 & O8 I(f) &  $-6.28$	& $386$\\
\hline
\end{tabular}
\tablefoot{
\tablefoottext{a}{In some figures we use the short notation between brackets.}
\tablefoottext{b}{$V$-magnitudes from (3),(4), and (5).}
\tablefoottext{c}{The value of $-6.35$ listed in Paper I is a typo. The correct value was used in the analysis.}}
\tablebib{
  (1) Paper I; (2) This work; (3) \citet{bresolin2007}; (4) \citet{bresolin2006}; (5) \citet{evans2007}.
}
\end{table*}

\par In the next section we give an overview of the observations and the data reduction. In Section~\ref{sec:analysis} we describe the analysis and present the results. We discuss the low-metallicity effective temperature scale in Section~\ref{sec:teff}, and the wind strengths in Section~\ref{sec:mdot}. Finally, we discuss the evolutionary properties of the sample and the recent star formation history of the host galaxies in Section~\ref{sec:discussionproperties}. We summarize our findings in Section~\ref{sec:summary}.

\section{Observations and data reduction}\label{sec:obs}

All stars have been observed with X-Shooter \citep{vernet2011} at ESO's Very Large Telescope as part of the NOVA program for guaranteed time observations.  An overview of the observational properties of the stars is given in Table~\ref{tab:stars}. Throughout this paper, we will use the identification given in this table.
\par The observations and data reduction of IC1613-1 to 3, WLM-1 and NGC3109-1 (program ID 085.D-0741) are described in Paper~I. An overview of observations of the other stars (under program IDs 088.D-0181 and 090.D-0212) is given in Table~\ref{tab:observations}. All stars were observed with a slit width of 0.8", 0.9" and 0.9" in the UVB, VIS and NIR arms, respectively. The corresponding resolving power $R$ is 6\,200 (UVB), 7\,450 (VIS) and 5\,300 (NIR). All observations were carried out while the moon was below the horizon or illuminated less than 30\% (dark conditions).

The data reduction of the newly observed stars was performed with the X-Shooter pipeline v2.2.0. To obtain uncontaminated 1D spectra, the science reduction was done without sky subtraction for each individual exposure. The resulting 2D spectra were folded in the wavelength direction and inspected for the presence of other objects in the slit. A clean part of the slit was then used for sky subtraction. The 1D spectra were extracted from the sky-subtracted 2D spectra. As the observed spectra suffer from nebular emission in the hydrogen and \ion{He}{I} lines, the extracted spectra were carefully inspected for residuals of nebular lines. If needed, a more suitable part of the slit was used. Whenever residuals remained after this procedure, they were clipped from the spectrum before the analysis.

\par The 1D spectra of the individual exposures were combined by taking the median flux at each wavelength so cosmic ray hits are removed. Finally, the extracted 1D spectra were normalized by fitting a 4th degree polynomial to the continuum, and dividing the flux by this function. Figure~\ref{fig:atlas} shows the resulting normalized spectra of all stars.

\par The spectra have a signal-to-noise ratio (S/N) between 25 and 45 per wavelength bin of 0.2 \AA \ in the UVB\footnote{Note that we are oversampling the spectral resolution. A S/N of 25 per wavelength bin of 0.2 \AA \ corresponds to a S/N of $\simeq$ 50 per resolution element at 4500 \AA. }. As expected in O stars, all the spectra show strong hydrogen, \ion{He}{i} and \ion{He}{ii} lines. Some spectra also show weak nitrogen lines. 

\section{Analysis \& results}\label{sec:analysis}
To investigate the properties of the target stars, we first obtained the stellar and wind parameters by fitting synthetic spectra to the observed line profiles. The method is described in the following section, and the results are presented in Section~\ref{sec:GAresults}. The stellar parameters were then used to obtain estimates of the evolutionary parameters (Section~\ref{sec:bonnsai}). We comment on the results of the individual targets in Section\ref{sec:individual}.

\begin{table}
\centering
\caption{Journal of observations.}\label{tab:observations}
\begin{tabular}{l l  c c c}
\hline\hline
ID & HJD & $t_{\mathrm{exp}}$ & \multicolumn{2}{c}{Average seeing} \\
 & {\tiny \textit{At start of obs.}}&  (s) & UVB (")& VIS (") \\
\hline \\[-8pt]
IC1613-5 & 2\,455\,858.653 & 4x900 & 1.1 & 1.0 \\
& 2\,455\,858.705 & 2x900 & 1.1 & 0.9 \\
NGC3109-2 & 2\,456\,337.531 & 4x900 & 2.2 & 1.2\\
& 2\,456\,337.587 & 4x900 & 1.4 & 1.0\\
 & 2\,456\,338.583 & 4x900 & 1.4& 1.0\\
 & 2\,456\,338.638 & 4x1200 & 1.0 & 0.8\\
NGC3109-3 & 2\,456\,337.643 & 4x1100 & 1.1& 0.8\\
& 2\,456\,337.732 & 4x1100 & 0.9 & 0.6\\
& 2\,456\,338.736 & 4x900 & 0.8 & 0.6\\
& 2\,456\,338.791 & 2x900 & 1.0 & 0.6\\
NGC3109-4 & 2\,456\,337.797 & 4x900 & 1.2& 0.7\\
& 2\,456\,337.853 & 2x1200 & 1.7 & 0.7\\
& 2\,456\,338.823 & 4x1200 & 1.5 & 0.7\\
\hline
\end{tabular}
\end{table}

\subsection{Fitting method}\label{sec:GA}
\begin{figure*}
   \resizebox{\hsize}{!}{\includegraphics{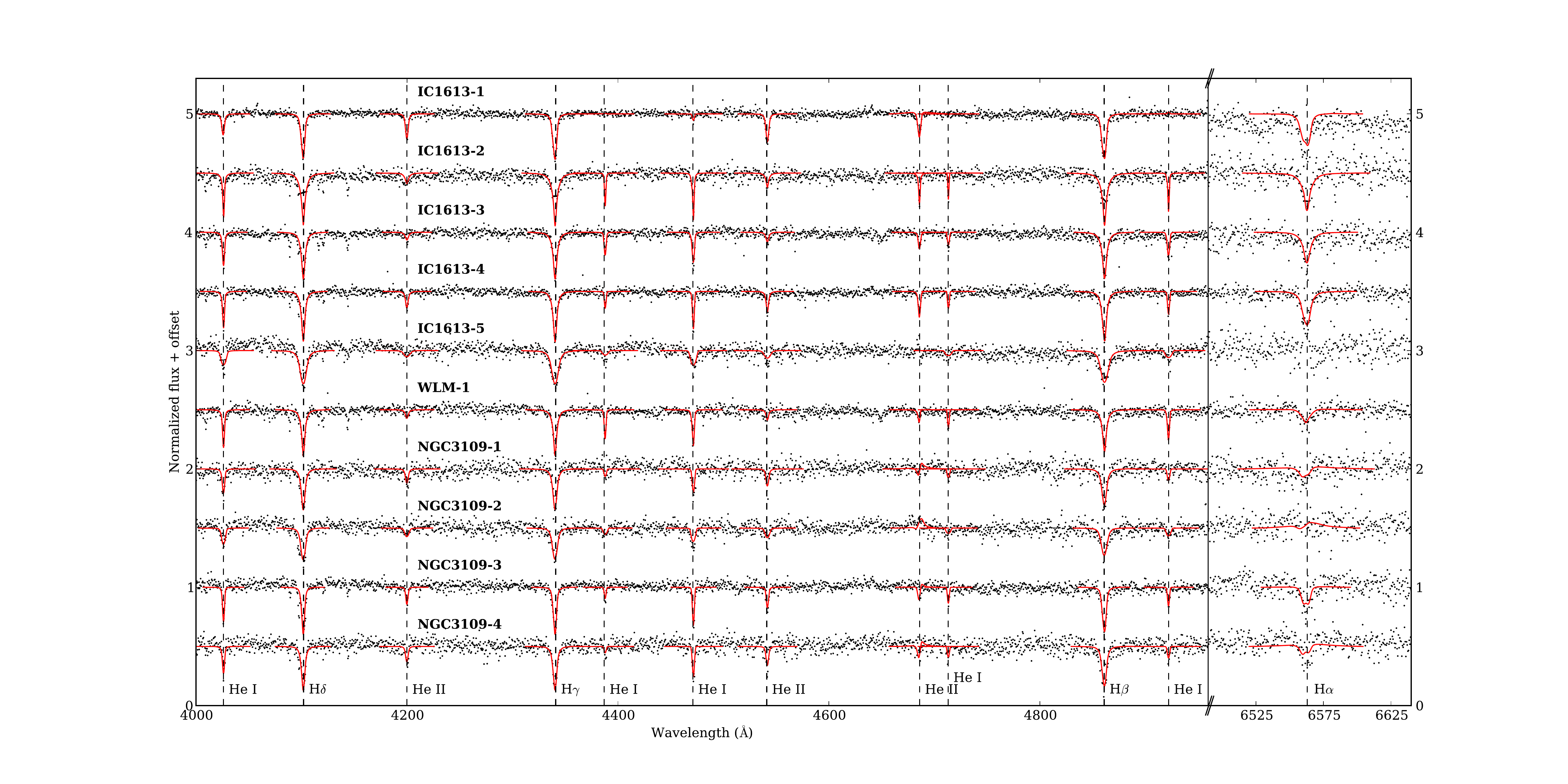}}
  \caption{Observed spectra (black dots) and best-fit line profiles (red lines). Rest wavelengths of the fitted spectral lines are indicated by the vertical dashed lines. In this plot the wavelength has been corrected for the radial velocities listed in Table~\ref{tab:stars}, and binned to 0.5 \AA.}
  \label{fig:atlas}
\end{figure*}

To determine the stellar and wind properties, we used an automated fitting method developed by \cite{mokiem2005}. This method fits spectra produced by the non-LTE model atmosphere code {\sc fastwind} \citep{puls2005} to the observed spectrum using the genetic algorithm based fitting routine {\sc pikaia} \citep{charbonneau1995}. This genetic algorithm (GA) method allows for a thorough exploration of parameter space in affordable CPU time on a supercomputer.
\par The absolute $V$-band magnitude $(M_\mathrm{V})$ is needed as input for the GA in order to determine the luminosity (Table~\ref{tab:stars}). $M_V$ was calculated using the $V$ magnitudes also given in Table~\ref{tab:stars} and distances and mean reddening listed in Table~\ref{tab:galaxies}. 
\par The radial velocity (RV) of each star is also listed in Table~\ref{tab:stars}. These were measured by fitting Gaussians to the \ion{H}{$\gamma$} and \ion{He}{i} $\lambda$4471 lines and calculating the average velocity needed to match the observed wavelength shifts.
\par The parameters that are obtained from the atmosphere fitting are the effective temperature $(T_{\mathrm{eff}})$, the surface gravity $(g)$, the mass-loss rate $(\dot{M})$, the surface helium abundance $(N_{\mathrm{He}})$, the atmospheric microturbulent velocity $(v_{\mathrm{tur}})$ and the projected rotational velocity $(v_{\mathrm{rot}}\sin{i})$. As in Paper I, the parameter describing the rate of acceleration of the outflow $(\beta)$ can not be constrained from the data, and was fixed to the value predicted by theory \citep[$\beta=0.95$ for the supergiants presented in this work;][]{muijres2012}. 
\par The terminal wind velocity $(v_{\infty})$ can not be constrained from the optical spectrum. Therefore, we used the empirical scaling with the escape velocity ($v_{\mathrm{esc}}$) for Galactic stars \citep[$v_{\infty}=2.65\,v_{\mathrm{esc}}$;][]{kudritzki2000}, and scaled these using \citet[][$v_{\infty} \propto Z^{0.13}$]{leitherer1992} to correct for the lower metallicity. In contrast to Paper I, this metallicity scaling has now been implemented into the GA, and was applied before running each individual {\sc fastwind} model. Therefore, the mass-loss rate no longer has to be scaled down after the fitting, as was previously needed.
\par Several additional small changes were implemented in the GA:
\begin{itemize}
\item{The best-fitting model is now selected based on the $\chi^2$, which is also used for the error calculation.}
\item{\ion{He}{i} $\lambda$4922 is now also fitted in addition to the 11 lines that were used in Paper I.}
\item{The minimum microturbulent velocity $v_{\mathrm{tur}}$ is set to 5 instead of 0, as {\sc fastwind} models with $v_{\mathrm{tur}} < 5$ may not be accurate.}
\item{The error on the flux is now based on the signal-to-noise ratio (S/N) calculated near each of the fitted lines, instead of a single value for each X-Shooter arm.}
\end{itemize}
To present a homogeneous analysis, we have re-analysed the stars of Paper I with the updated fitting routine, and we present the new parameters here. In general, the new values are in excellent agreement with those presented in Paper I. The exception is IC1613-A1, where \ion{H}{$\alpha$} was not properly normalized in Paper I. For this star we re-normalized \ion{H}{$\alpha$}, resulting in a somewhat higher mass-loss rate.
\par Table~\ref{tab:bestfit} presents the best-fit parameters for each of the target stars. The synthetic spectra of the corresponding {\sc fastwind} models are overplotted on the observed spectra in Figure~\ref{fig:atlas}. 

\subsection{Derived properties and error calculation}\label{sec:GAresults}

\begin{table*}
\centering
\caption{Best-fitting stellar and wind parameters.}\label{tab:bestfit}
\begin{tabular}{l c c c c c c}
\hline\hline
ID & $T_{\mathrm{eff}}$ & $\log{g}$ & $\log{\dot{M}}$ & $N_{\mathrm{He}}/N_{\mathrm{H}}$ & $v_{\mathrm{tur}}$ & $v_\mathrm{rot} \sin{i}$ \\
& \tiny{(kK)} & \tiny{(cm s$^{-2}$)} & \tiny{($M_{\odot}$ yr$^{-1}$)} & & \tiny{(km s$^{-1}$)} &\tiny{(km s$^{-1}$)} \\
\hline \\[-8pt]
IC1613-1		& $45.40^{+2.00}_{-2.25}$	& $3.65^{+0.16}_{-0.10}$	 	& $-5.85^{+0.10}_{-0.50}$		& $0.25^{\uparrow}_{0.10}$		& $24^{+6}_{\downarrow}$	& $98^{+36}_{-44}$ \\
IC1613-2		& $33.85^{+2.10}_{-2.75}$	& $3.77^{+0.33}_{-0.33}$		& $-6.35^{+0.35}_{\downarrow}$	& $0.14^{+0.15}_{-0.08}$		& $11^{+15}_{\downarrow}$	& $32^{+38}_{-22}$ \\
IC1613-3		& $31.45^{+1.65}_{-2.45}$	& $3.41^{+0.22}_{-0.19}$		& $-6.25^{+0.35}_{-1.20}$		& $0.15^{+0.09}_{-0.08}$		& $5^{+18}_{\downarrow}$	& $94^{+32}_{-24}$	\\
IC1613-4 		& $35.2^{+1.85}_{-1.40}$		& $3.52^{+0.20}_{-0.11}$		& $-6.25^{+0.15}_{-0.50}$		& $0.12^{+0.10}_{-0.04}$		& $17^{+7}_{\downarrow}$	& $76^{+16}_{-26}$	\\
IC1613-5		& $35.05^{+4.55}_{-4.8}$ & $3.74^{\uparrow}_{-0.44}$	& $-$						& $0.06^{+0.17}_{\downarrow}$		& $27^{\uparrow}_{\downarrow}$				& $270^{+112}_{-92}$\\
WLM-1		& $30.60^{+1.70}_{-3.60}$	& $3.28^{+0.19}_{-0.31}$		& $-5.50^{+0.10}_{-0.35}$		& $0.22^{\uparrow}_{0.10}$ & $9^{+9}_{\downarrow}$	& $72^{+36}_{-22}$	\\
NGC3109-1	& $35.15^{+3.20}_{-2.55}$	& $3.53^{+0.30}_{-0.42}$		& $-5.35^{+0.15}_{-0.35}$		& $0.09^{0.22}_{0.04}$ & $17^{\uparrow}_{\downarrow}$	&  $110^{+50}_{-52}$	\\
NGC3109-2	& $33.30^{+3.30}_{-2.25}$ & $3.35^{+0.43}_{-0.19}$ & $-5.45^{+0.30}_{-0.15}$ 		& $0.09^{+0.18}_{-0.05}$		& $10^{\uparrow}_{\downarrow}$	& $200^{+102}_{-104}$ \\
NGC3109-3	& $33.05^{+1.45}_{-1.25}$ & $3.16^{+0.19}_{-0.11}$ & $-5.75^{+0.15}_{-0.25}$		& $0.12^{+0.17}_{-0.03}$		& $23^{+3}_{\downarrow}$			& $82^{+30}_{-32}$\\
NGC3109-4	& $35.05^{+3.45}_{-4.45}$ & $3.43^{+0.51}_{-0.31}$		& $-5.55^{+0.25}_{-0.45}$		& $0.08^{+0.15}_{-0.04}$		& $25^{+5}_{\downarrow}$	& $96^{+60}_{-70}$\\
\hline
\end{tabular}
\end{table*}

\begin{table*}
\centering
\caption{Properties derived from best-fit parameters.}\label{tab:derivedproperties}
\begin{tabular}{l c c c c c}
\hline\hline
ID &$v _{\infty}$ & $\log{L}$  & $R$ & $M_{\mathrm{spec}}$ & $\log{D_\mathrm{mom}}$ \\
& \tiny{(km\,s$^{-1}$)} & \tiny{($L_{\odot}$)} & \tiny{($R_{\odot}$)} & \tiny{($M_{\odot}$)} & \tiny{(g\,cm\,s$^{-2}$\,$R_{\odot}^{1/2}$)}\\
\hline \\[-8pt]
IC1613-1		& $1755^{+328}_{-165}$	& $5.71^{+0.05}_{-0.06}$	& $11.9^{+0.4}_{-0.3}$ 	& $22.6^{+8.4}_{-3.6}$ & $28.73^{+0.11}_{-0.53}$ \\
IC1613-2		& $2022^{+872}_{-593}$	& $5.21^{+0.06}_{-0.09}$	& $11.9^{+0.8}_{-0.5}$	& $30.2^{+29.0}_{-14.5}$ & $28.29^{+0.46}_{\downarrow}$ \\
IC1613-3		& $1625^{+440}_{-296}$	& $5.42^{+0.06}_{-0.08}$	& $17.6^{+1.2}_{-0.7}$	& $28.9^{+16.4}_{-8.9}$ & $28.38^{+0.42}_{-1.19}$ \\
IC1613-4		& $1558^{+387}_{-184}$	& $5.32^{+0.06}_{-0.04}$	& $12.5^{+0.4}_{-0.5}$	& $18.9^{+10.1}_{-4.2}$ & $28.29^{+0.19}_{-0.51}$ \\
IC1613-5		& $2010^{+2672}_{-742}$ & $5.32^{+0.13}_{-0.16}$ & $12.6^{+1.5}_{-1.1}$ & $31.6^{+129.8}_{-18.4}$ & $-$\\
WLM-1		& $1777^{+435}_{-488}$	& $5.79^{+0.06}_{-0.13}$	& $28.3^{+3.0}_{-1.2}$	& $55.6^{+30.5}_{-24.3}$ & $29.28^{+0.14}_{-0.45}$ \\
NGC3109-1		& $2166^{+876}_{-1374}$	& $5.87^{+0.10}_{-0.08}$	& $23.7^{+1.4}_{-1.5}$	& $69.1^{+65.3}_{-39.7}$ & $29.47^{+0.28}_{-0.53}$ \\
NGC3109-2		& $1692^{+1058}_{-331}$ & $5.71^{+0.10}_{-0.13}$ & $21.9^{+1.2}_{-1.5}$ & $38.9^{+65.9}_{-13.7}$ & $29.25^{+0.51}_{-0.24}$\\
NGC3109-3		& $1357^{+328}_{-156}$& $5.69^{+0.05}_{-0.04}$ & $21.8^{+0.7}_{-0.7}$ & $25.0^{+13.3}_{-5.32}$ & $28.85^{+0.24}_{-0.26}$\\
NGC3109-4		& $1767^{+1412}_{-492}$ & $5.71^{+0.11}_{-0.14}$ & $19.8^{+0.5}_{-1.3}$ 	& $38.5^{+86.0}_{-17.2}$ & $29.15^{+0.45}_{-0.56}$\\
\hline
\end{tabular}
\end{table*}

\par In addition to $v_{\infty}$ (discussed above), several important quantities can be derived from the best-fit parameters: the bolometric luminosity ($L$), the stellar radius ($R$), and the spectroscopic mass ($M_{\mathrm{spec}}$). These are given in Table~\ref{tab:derivedproperties}. This table also gives the modified wind momentum, which is defined as $D_{\mathrm{mom}} = \dot{M} \, v_{\infty} \, \sqrt{R/R_{\odot}}$ and is ideal to study the mass loss as it is almost independent of mass. Furthermore, $D_{\mathrm{mom}}$ scales with the luminosity through $R$, making it less sensitive to uncertainties in the luminosity determination.
\par To derive the error bars given in Tables~\ref{tab:bestfit} and \ref{tab:derivedproperties}, we first divide all $\chi^2$ values with a factor such that the best model has a reduced $\chi^2$ of unity. This ensures meaningful error bars that are not influenced by under or overestimated errors on the flux (see Paper I). We then calculate the probability $P = 1 - \Gamma(\chi^2/2, \nu/2)$, with $\Gamma$ the incomplete gamma function and $\nu$ the degrees of freedom, for each model. $P$ quantifies the probability that a $\chi^2$ value differs from the best-fit $\chi^2$ due to random fluctuations. Models that satisfy $P \geq 0.05$ are accepted as providing a suitable fit, and the range covered by the stellar parameters of these models (and the properties derived from them) is taken as the 95\% confidence interval. This method also ensures that uncertainties in the parameters that arise due to clipped parts of the spectrum are reflected in the error bars.

\subsection{Mass and age}\label{sec:bonnsai}
To determine the evolutionary parameters of the stars, we use the {\sc bonnsai}\footnote{The {\sc bonnsai} web-service is available at bonnsai.astro.uni-bonn.de.} tool (Schneider et al. in prep.). {\sc bonnsai} uses Bayes' theorem to constrain key stellar parameters, such as initial mass and age, by comparing the observed stellar parameters to theoretical predictions from stellar evolution. 
\par We obtained an estimate of the initial mass $(M_{\mathrm{ini}})$, the current mass $(M_{\mathrm{act}})$, the initial and current rotation ($v_{\mathrm{rot,ini}}$ and $v_{\mathrm{rot,act}}$), and the age of the stars. We use the evolutionary tracks for SMC metallicity of \cite{brott2011}, as these are closest in metallicity. As we do not find a significant difference in the temperature of our stars compared to similar SMC stars (see Section~\ref{sec:teff}), the use of the SMC tracks does not induce large systemetic uncertainties in the evolutionary parameters. As priors to the {\sc bonnsai} method we choose a \cite{salpeter1955} initial mass function, and the \cite{ramirez2013} 30 Doradus distribution for the initial rotational velocity.
\par As input observables we used the luminosity, effective temperature, surface gravity and projected rotational velocity. {\sc bonnsai} adapts these parameters based on the comparison with the evolutionary predictions. The posterior reproduced parameters are within errors of the input values. The estimated evolutionary parameters are given in Table~\ref{tab:bonnsai}. The stellar masses that are derived with the {\sc bonnsai} method are in good agreement with the mass estimates that would be derived using the conventional method, i.e. a visual comparison with evolutionary tracks in the Herzsprung-Russell diagram (HRD, Figure~\ref{fig:hrd}).


\begin{table*}
\centering
\caption{Stellar parameters obtained from comparison with evolutionary tracks using {\sc Bonnsai}}.\label{tab:bonnsai}
\begin{tabular}{l c c c c c}
\hline\hline
ID 			& $M_{\mathrm{ini}}$	& $M_{\mathrm{act}}$	& $v_{\mathrm{rot,ini}}$	& $v_{\mathrm{rot, act}}$	& $\tau$ \\
			& \tiny{($M_{\odot}$)} 	& \tiny{($M_{\odot}$)}	& \tiny{(km\,s$^{-1}$)}	& \tiny{(km\,s$^{-1}$)}		& \tiny{(Myr)} \\
\hline \\[-8pt]
IC1613-1		& $49.0^{+3.5}_{-3.4}$	& $47.6^{+3.6}_{-3.1}$	& $100^{+48}_{-40}$	& $100^{+80}_{-54}$	& $2.32^{+0.31}_{-0.34}$	\\
IC1613-2		& $24.6^{+2.1}_{-1.9}$	& $24.4^{+1.9}_{-1.8}$	& $70^{+41}_{-38}$		& $70^{+41}_{-38}$		& $5.00^{+0.71}_{-0.62}$	\\
IC1613-3		& $29.4^{+2.4}_{-2.3}$	& $28.8^{+2.2}_{-2.2}$	& $110^{+40}_{-36}$	& $100^{+49}_{-27}$	& $4.74^{+0.45}_{-0.36}$	\\
IC1613-4		& $28.8^{+1.8}_{-1.4}$	& $28.4^{+1.7}_{-1.3}$	& $80^{+37}_{-28}$		& $80^{+37}_{-28}$		& $4.40^{+0.32}_{-0.36}$	\\
WLM-1		& $41.6^{+6.4}_{-5.3}$	& $39.8^{+6.1}_{-4.7}$	& $90^{+48}_{-29}$		& $90^{+45}_{-32}$		& $3.58^{+0.49}_{-0.33}$	\\
NGC3109-1	& $52.6^{+5.1}_{-4.3}$	& $50.0^{+5.1}_{-3.6}$	& $110^{+59}_{-45}$	& $110^{+59}_{-46}$	& $2.86^{+0.29}_{-0.22}$	\\
NGC3109-2	& $40.0^{+6.4}_{-5.4}$	& $38.6^{+6.0}_{-5.0}$	& $130^{+111}_{-61}$	& $130^{+113}_{-63}$	& $3.44^{+0.50}_{-0.41}$	\\
NGC3109-3	& $42.0^{+2.7}_{-2.3}$	& $40.4^{+2.6}_{-2.0}$	& $90^{+48}_{-31}$		& $90^{+48}_{-31}$		& $3.54^{+0.19}_{-0.17}$	\\
NGC3109-4	& $39.8^{+6.9}_{-5.9}$	& $38.6^{+6.5}_{-5.4}$	& $100^{+62}_{-47}$	& $100^{+63}_{-48}$	& $3.34^{+0.58}_{-0.47}$	\\
\hline
\end{tabular}
\end{table*}

\subsection{Comments on individual stars}\label{sec:individual}

\begin{table*}
\centering
\caption{Coefficients for the spectral type - $T_{\mathrm{eff}}$ calibrations.}\label{tab:SpT_Teff}
\begin{tabular}{l c c c c}
\hline\hline
 			& \multicolumn{2}{c}{Unweighted}	& \multicolumn{2}{c}{Weighted} \\
Sample & a & b & a & b \\
\hline\\[-8pt]
This work			& 47584.3		& $-$1595.4		& 44907.9$\pm$4457.2	& $-$1339.1$\pm$518.9	\\
Low Z			& 53398.0		& $-$2203.6		& 56636.1$\pm$766.9	& $-$2677.4$\pm$106.1	\\
Low Z (no O3)		& 48670.7		& $-$1653.2		& 47283.6$\pm$3017.7	& $-$1588.2$\pm$356.1	\\
SMC				& 51929.7		& $-$2138.8		& 50189.7$\pm$1329.2	& $-$1957.2$\pm$167.4	\\
\hline
\end{tabular}
\tablefoot{Here, $a$ and $b$ are the coefficients in $T_{\mathrm{eff}} = a + b \times X_{\mathrm{SpT}}$, with $X_{\mathrm{SpT}}$ the O subtype.}
\end{table*}

\subsubsection{IC1613-1}
This is the only dwarf star in the sample, which is reflected by its young derived age (Table~\ref{tab:bonnsai}). In Paper I, this was the only star with a wind momentum lower than the empirical SMC values from \cite{mokiem2007}. However, as already mentioned, \ion{H}{$\alpha$} was not properly normalized, which caused the mass-loss rate to be slightly underestimated. In our new analysis of this star we renormalized \ion{H}{$\alpha$}, and the updated modified wind momentum is now comparable to those found for SMC stars.
\par \cite{garcia2014} obtained the UV spectrum of IC1613-1 using HST-COS, and used it to determine the terminal wind velocity. They find $v_{\infty} = 2\,200^{+150}_{-100}$ km s$^{-1}$, somewhat higher than the values of 1\,869 km s$^{-1}$ (Paper I) and 1\,755 km s$^{-1}$ (this work) that we obtain from the scaling with the escape velocity.  Their value for the terminal wind velocity would result in a value of $\log(D_{\mathrm{mom}})$ that is $\simeq 0.1$ dex higher than ours. 
\par IC1613-1 has a low surface gravity for its luminosity class, and is enriched in helium.
 
 \subsubsection{IC1613-2}
 The nebular emission is variable along the X-shooter slit, which prevents a good nebular subtraction. As a consequence, a large part of the core of the Balmer lines had to be clipped from the spectrum before fitting. Without the core of \ion{H}{$\alpha$} we can only derive an upper limit for the mass-loss rate of this star.

\subsubsection{IC1613-3 and IC1613-4}
Both these stars are well fitted by the atmosphere models. For both stars \ion{H}{$\alpha$} is strongly in absorption and the mass-loss rate cannot be well constrained from this line. This results in fairly large error bars on their modified wind momenta.

\subsubsection{IC1613-5}
After our observations of this object, it was found to be an eclipsing binary \citep{bonanos2013}. Our spectra show strong variability in \ion{He}{ii} $\lambda$4686 and \ion{H}{$\alpha$} between individual exposures, which may be due to colliding winds \citep{stevens1992}. Although we provide parameters from fitting the other lines, it is likely that the spectrum is composite (depending on the mass ratio). The listed values are therefore only representative of the composite spectrum. This is a possible cause of the broad spectral lines of IC1613-5 (see Figure~\ref{fig:atlas}), although the rotational velocity of $v_{\mathrm{rot}}\sin{i} = 270$ km s$^{-1}$ that is needed to fit these lines is not unphysically high \citep[see][]{ramirez2013}. The variability does prevent us to constrain the mass-loss rate and consequently modified wind momentum of this star. We excluded this star from both the {\sc bonnsai} analysis and our discussion of the mass-loss rates (Section~\ref{sec:mdot}).

\subsubsection{WLM-1}
This is the only star in our sample in WLM, and one of the stars from Paper I which showed a large discrepancy with radiation-driven wind theory. The wind properties derived with the updated GA are very similar to those of Paper I. This star also has a high helium abundance.

\subsubsection{NGC3109-1}
This star shows a strong stellar wind, and is one of the stars in Paper I that exhibits the largest discrepancy with radiation-driven wind theory. While \ion{H}{$\alpha$} is still slightly in absorption, \ion{He}{ii} $\lambda$4686 is fully filled in.

\subsubsection{NGC3109-2 and NGC3109-4}
Both stars show signs of strong winds in their spectrum. In the fitting, the line center of \ion{H}{$\alpha$} in NGC3109-4 was clipped due to nebular contribution that could not be fully corrected for. \ion{He}{ii} $\lambda$4686 is fully filled in in both stars. \ion{H}{$\alpha$} is fully filled in for NGC3109-2 and almost fully filled in for NGC3109-3. Unsurprisingly, the derived wind momenta for both stars are high.

\subsubsection{NGC3109-3}
\ion{He}{ii} $\lambda$4686 is almost fully filled in, while \ion{H}{$\alpha$} is still mildly in absorption. This results in a modest wind strength.

\section{Effective temperature scale}\label{sec:teff}

\begin{figure}[!t]
   \resizebox{\hsize}{!}{\includegraphics{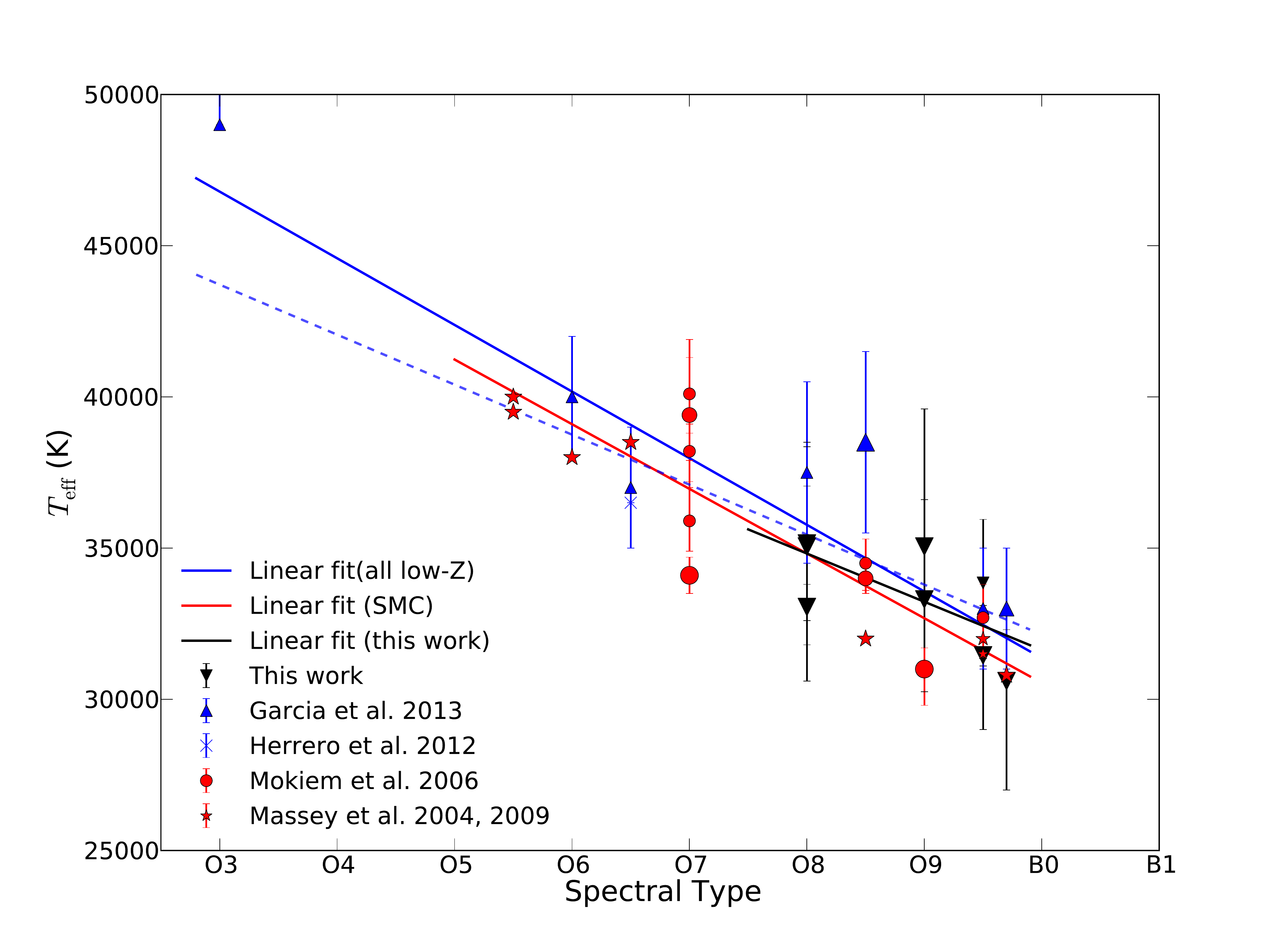}}
  \caption{Spectral type versus effective temperature calibration for giants and supergiants in low-metallicity environments. Symbol size indicates the luminosity class, with the larger symbols for supergiants. Plotted are results from this work, the \cite{herrero2012} and \cite{garcia2013}  results for IC1613, and the \cite{mokiem2006} and \cite{massey2004, massey2009} results for the SMC. The solid black line indicates a linear fit to the stars from this paper, not including the error bars on $T_{\mathrm{eff}}$. The red solid line is the unweighted linear fit to the SMC stars, and the blue solid line the fit to the stars in IC~1613,  WLM and NGC~3109 (low-Z). The dashed blue line is an unweighted fit to all low-Z results but excluding the single O3 giant. It illustrates the sensitivity of the found relation to this point.}
  \label{fig:teffcal}
\end{figure}

\cite{garcia2013} presented the first effective temperature calibration for potentially sub-SMC metallicities (their figure~7). In Figure~\ref{fig:teffcal} we use our results (Table~\ref{tab:bestfit}) to provide an updated version of this calibration. Similar to \cite{garcia2013}, we first determined the spectral type - $T_{\mathrm{eff}}$ relation using an unweighted least-squares linear fit to the temperatures of the giants and supergiants from our work. We did the same for the total sample of low-Z giants and supergiants \citep[this work;][]{herrero2012, garcia2013} and an SMC sample from \cite{mokiem2006} and \cite{massey2004, massey2009}. The coefficients of the derived linear relations are given in Table~\ref{tab:SpT_Teff}.
\par The updated $T_{\mathrm{eff}}$ scale for low metalicities is very similar to the relation found by \cite{garcia2013}, and is $\simeq 1000$ K hotter than the SMC relation. This is expected for a lower metallicity, as the stars are hotter due to a slightly smaller radius \citep[the result of lower opacities in the stellar interior; see e.g., ][]{mokiem2004}. However, as \cite{garcia2013} conclude, the significance of the observed difference between the temperature scales is unclear, given the error bars on the temperatures. Also, the low-Z relation is very sensitive to the position of the only O3 III star in IC~1613, in the region of parameter space where the SMC relation is not constrained. This sensitivity is illustrated by the dotted blue line in Figure~\ref{fig:teffcal}, which is the relation found when the O3 star is excluded from the fit. 

\begin{figure}[!t]
   \resizebox{\hsize}{!}{\includegraphics{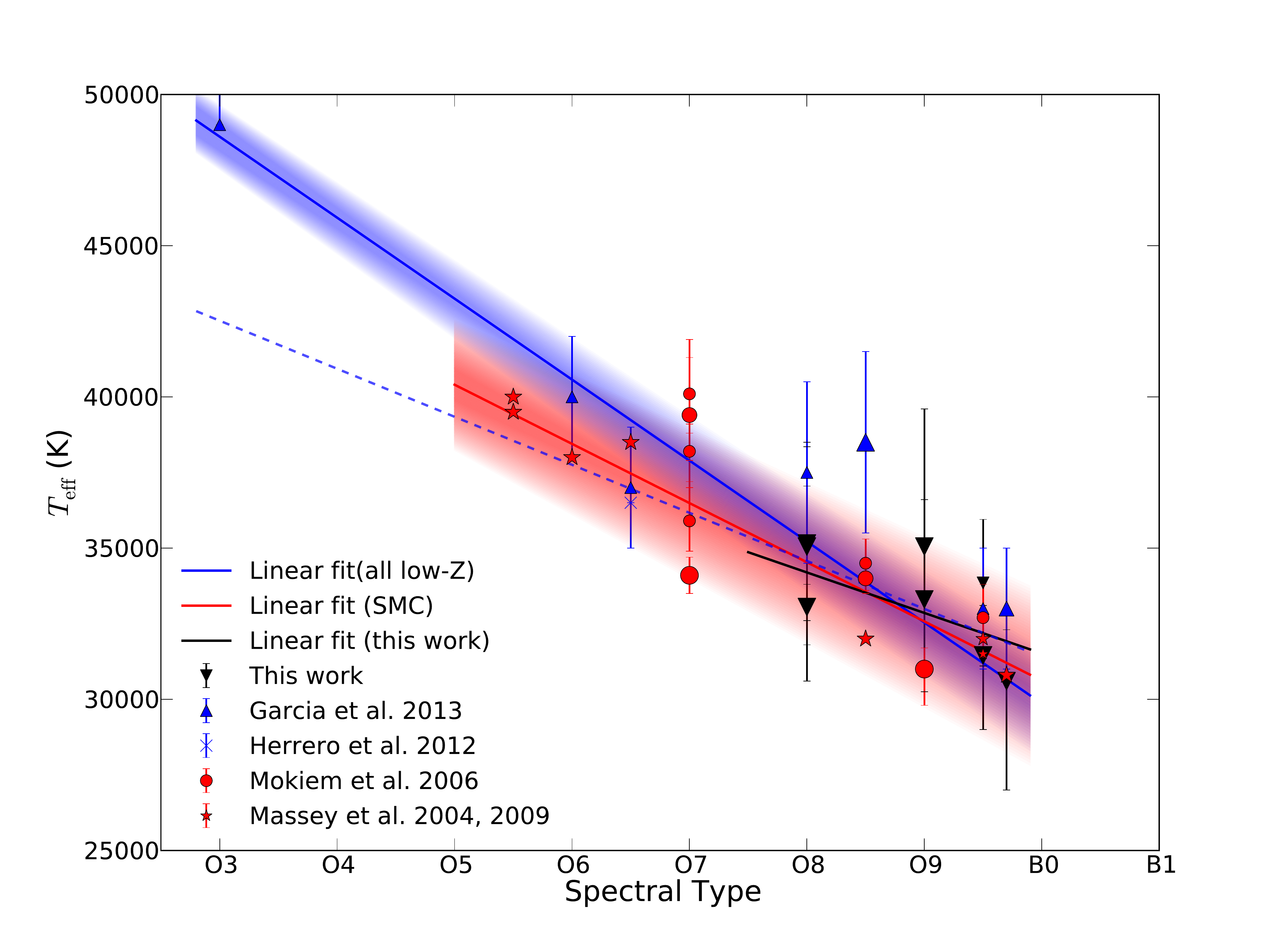}}
  \caption{Same as Figure~\ref{fig:teffcal}, but with the fitting done including the error bars on $T_{\mathrm{eff}}$. For stars that do not have published error bars, an error of 1 kK was adopted. The shaded areas indicate the uncertainties of the relations found for the SMC and low-metallicity relations.}
  \label{fig:teffcal_errors}
\end{figure}

\par As a second step, we included the error bars in our analysis. Figure~\ref{fig:teffcal_errors} presents the relations that are obtained by weighted least-squares linear fits (i.e. including the error bars on $T_{\mathrm{eff}}$) to the same samples. As the error bars on the temperature presented in Table~\ref{tab:bestfit} correspond to the 95\% confidence interval, we use half these values (roughly corresponding to $\sim1 \sigma$ for normally distributed errors). Because symmetric error bars are easier to handle in a simple approach, we use the average of the upper and lower errors. \cite{massey2004, massey2009} do not provide error bars, and we adopt $\pm$1\,000 K for these stars. The coefficients of the relations that we obtain are given in Table~\ref{tab:SpT_Teff}.
\par The low-metallicity temperature scale obtained from the weighted fits is steeper than the SMC scale, and no longer above it at each point of the spectral range. The error bars on both relations overlap over the entire spectral range covered by the SMC stars. Additionally, the effective temperature at spectral type O8 obtained from our sample, which is well constrained by four stars, is very close to the SMC value regardless of the fitting method. 
\par Thus, with the number of stars that are currently analysed we do not find a significant difference between the effective temperature calibrations for the host galaxies of the stars studied in this paper and the SMC. However, a good comparison is hampered by the small sample size, and the absence of early-type giants and supergiants in the SMC sample. Ideally, the low-metallicity effective temperature scale has to be derived from a large number of dwarfs of all subtypes, which are also found in the SMC. However, even if a sufficient number of O-type dwarfs is present in the low-metallicity galaxies, obtaining their spectra will have to await the advent of 30m-class telescopes.

\section{Mass loss versus metallicity}\label{sec:mdot}

\begin{figure}[!t]
   \resizebox{\hsize}{!}{\includegraphics{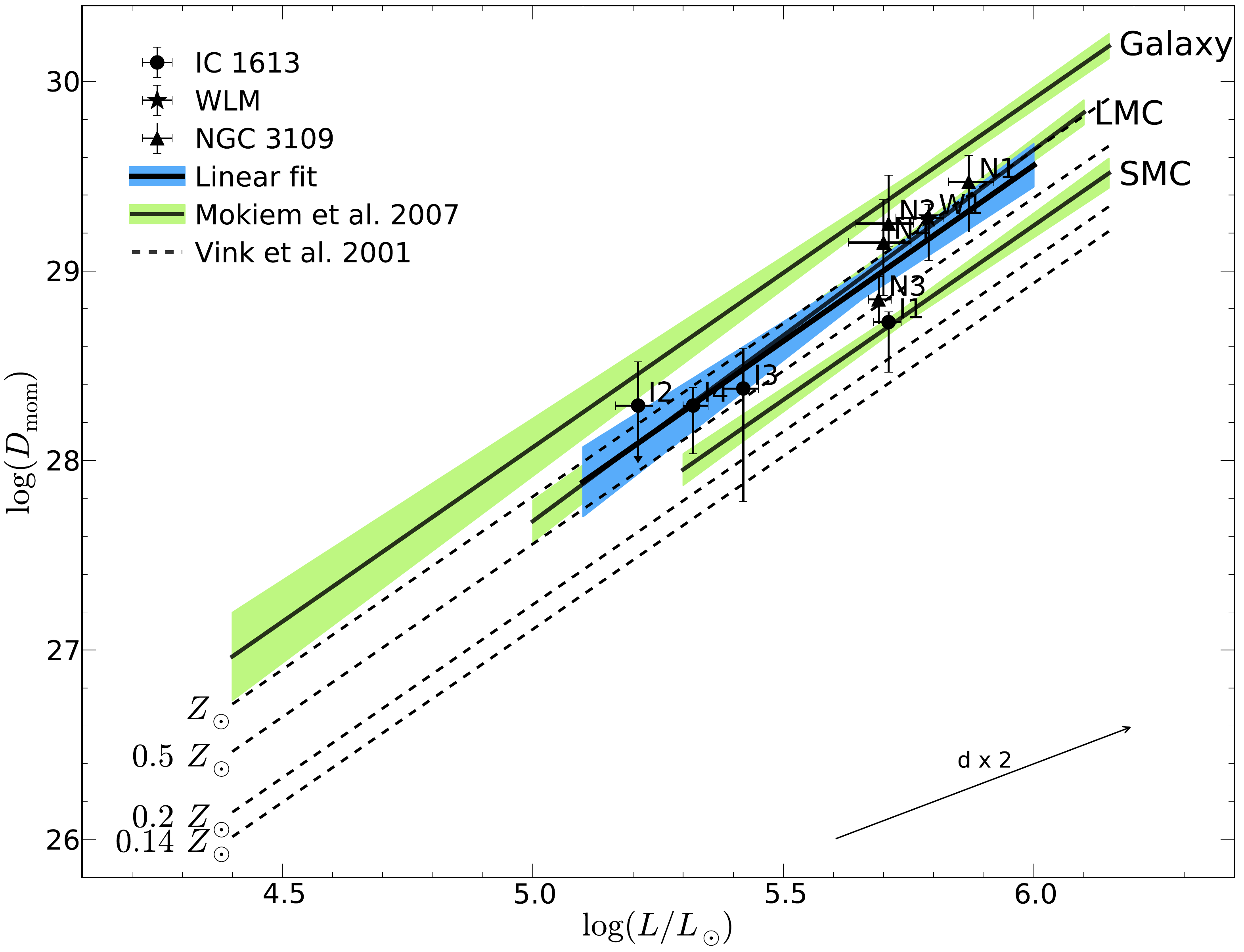}}
  \caption{Location of the target stars in the modified wind momentum versus luminosity diagram. Also indicated are the theoretical predictions from \cite{vink2001} and the empirical results from \cite{mokiem2007}. NGC3109-4 is shifted by $-$0.01 dex in luminosity for clarity. $1 \sigma$ error bars are indicated. The thick line represents a linear fit to our results, and shows that the wind strengths are comparable to the empirical LMC results.}
  \label{fig:wld}
\end{figure}

In Paper I, we reported that the wind momenta of the stars in our sample appear to be higher than theoretically predicted for their metallicity. This trend remains after refitting these stars with the updated GA and including the new targets. This is shown in the updated modified wind momentum versus luminosity diagram (WLD; Figure~\ref{fig:wld}). This figure also shows a weighted linear fit to our data ($\log{D_{\mathrm{mom}}} = a + b \times \log{L/L_{\odot}}$, with $a=18.4 \pm 1.9$, $b=1.86 \pm 0.33$). The fit confirms that the stars exhibit LMC strength winds. Only two stars (IC1613-1 and NGC3109-3) have a best-fit value for their modified wind momentum that is close to SMC values, and none have values indicative of a sub-SMC wind strength.
\par An important aspect to note when using the WLD to compare mass-loss rates, is that inhomogeneities in the wind (clumping) are not taken into account when deriving the empirical mass-loss rate. This neglect of clumping causes mass-loss rates derived from diagnostic lines sensitive to the density-squared, such as \ion{He}{ii} $\lambda$4686 and \ion{H}{$\alpha$}, to be over-estimated \citep[e.g.,][]{puls2008}.
\par This effect can be seen in Figure~\ref{fig:wld} by comparing the results from \cite{mokiem2007} to the predictions from \cite{vink2001}. The empirical values for the Galaxy, LMC and SMC are clearly higher than the ones predicted by theory. However, the trend of decreasing wind strength at lower metallicities is in excellent agreement with theory. Thus, for an assumed sub-SMC metallicity we would expect our stars to be located below the empirical SMC values in the WLD.
\par \cite{lucy2012} argues that the neglect of wind clumping is the most likely explanation for the high mass-loss rates of these stars. However, we do argue that wind clumping would have to be metallicity dependant to explain our results. An other possibility given by \cite{lucy2012} is the presence of an additional wind-driving mechanism, possibly only operating in winds that have low terminal wind velocities, or in a restricted part of $(T_{\mathrm{eff}}, g, Z)$-space. 
\par However, an explanation for the high mass-loss rates may be found in the assumed metallicity of the host galaxies. Iron (and iron-like elements) remains the dominant element in driving the wind for metallicities down to $0.1 Z_{\odot}$, while $\alpha$ elements dominate at lower metallicities \citep{vink2001}. Thus, the iron content of the stars needs to be evaluated to be able to properly compare the wind strengths with theoretical predictions.
\par While all the host galaxies have very low average stellar iron abundances of $[\mathrm{Fe}/\mathrm{H}]\la -1.2$ \citep[for an overview, see][]{mcconnachie2012}, the metallicity of the young stellar population is likely higher. This metallicity can be constrained indirectly from \ion{H}{ii} regions and directly from red and blue supergiants. \cite{garcia2014}, \cite{levesque2012}, and \cite{evans2007} give overviews of all relevant metallicity measurements of the young stellar population of IC~1613, WLM, and NGC3109, respectively. The metalicity measurements range from $0.05\, Z_{\odot}$ to $0.10\, Z_{\odot}$ based on the oxygen abundance in \ion{H}{ii} regions and up to $0.15\, Z_{\odot}$ in blue supergiants. However, there are indications that the galaxies have a sub-solar $\alpha$-to-iron ratio. Below, we give an overview of the {\it stellar} iron (or iron-group elements) abundance measurements in the young population of the host galaxies.
\par \cite{tautvaisiene2007} derive an iron content of $[\mathrm{Fe}/\mathrm{H}] = -0.67\pm0.06$ for three M-type supergiants in IC~1613, or $Z_{\mathrm{IC1613}}=0.21\,Z_{\odot}$ based on iron. \cite{garcia2014} find qualitative indications that the iron content might be close to the SMC value. 
\par \cite{venn2003} report an iron abundance of $[\mathrm{Fe}/\mathrm{H}] = -0.38\pm0.29$ for two supergiants in WLM, corresponding to $Z_{\mathrm{WLM}}=0.42\,Z_{\odot}$, but with very large error bars. They derive a stellar oxygen abundance that is five times higher than those found from nebular studies. Conversely, \cite{urbaneja2008} derive $Z_{\mathrm{WLM}}=0.13\,Z_{\odot}$ based on mainly iron, chromium and titanium in blue supergiants, and find no indication that the $\alpha$-to-iron ratio is non-solar. In particular, they derive a metallicity of $[Z] = -0.80 \pm 0.20$ for WLM-1 ($0.16\,Z_{\odot}$). It therefore seems unlikely that an underestimated iron abundance explains the strong stellar wind of WLM-1.
\par \cite{hosek2014} analysed 12 late-B and early-A supergiants in NGC3109, and derive $[Z] = -0.67 +/- 0.13$ based on iron-group elements, or $Z_{\mathrm{NGC3109}}=0.21\pm0.08\,Z_{\odot}$. As for IC~1613, this indicates that the iron content is SMC-like. Our results should thus be compared to the SMC predictions.
\par For our sample stars, an SMC metallicity would lessen the discrepancy between the observed wind momenta and those predicted from theory. Compared to the SMC predictions, IC1613-1 is in good agreement with the radiation-driven wind theory, while the other three stars in IC1613 have too high best-fit values but agree within error bars. NGC3109-3 has a slightly too high mass-loss rate but is in agreement within errors. For the other three stars in NGC3109 the best-fit wind strengths are comparable to or slightly higher than LMC values, but can just be reconciled with SMC values within errors for two of them. The wind strength of WLM-1 is just in agreement with an SMC metallicity, but as mentioned above, it is unlikely that the metallicity is underestimated for this star. 
\par Considering the sample as a whole, the observed discrepancy with radiation-driven wind theory at low metallicities may be reduced if the metallicity has indeed been underestimated. However, the stars still tend to have too high mass-loss rates, even if their iron content is comparable to SMC stars. For our results to be fully in agreement with the predictions from radiation-driven wind theory, the iron content should be LMC-like, or half solar (see Figure~\ref{fig:wld}). Further constraints of the metal content in the host galaxies would be helpful. Most importantly, a confirmation of the wind properties from the UV has to be obtained to reduce the uncertainties in the derived wind momenta.

\section{Evolutionary state }\label{sec:discussionproperties}

\begin{figure}[!t]
   \resizebox{\hsize}{!}{\includegraphics{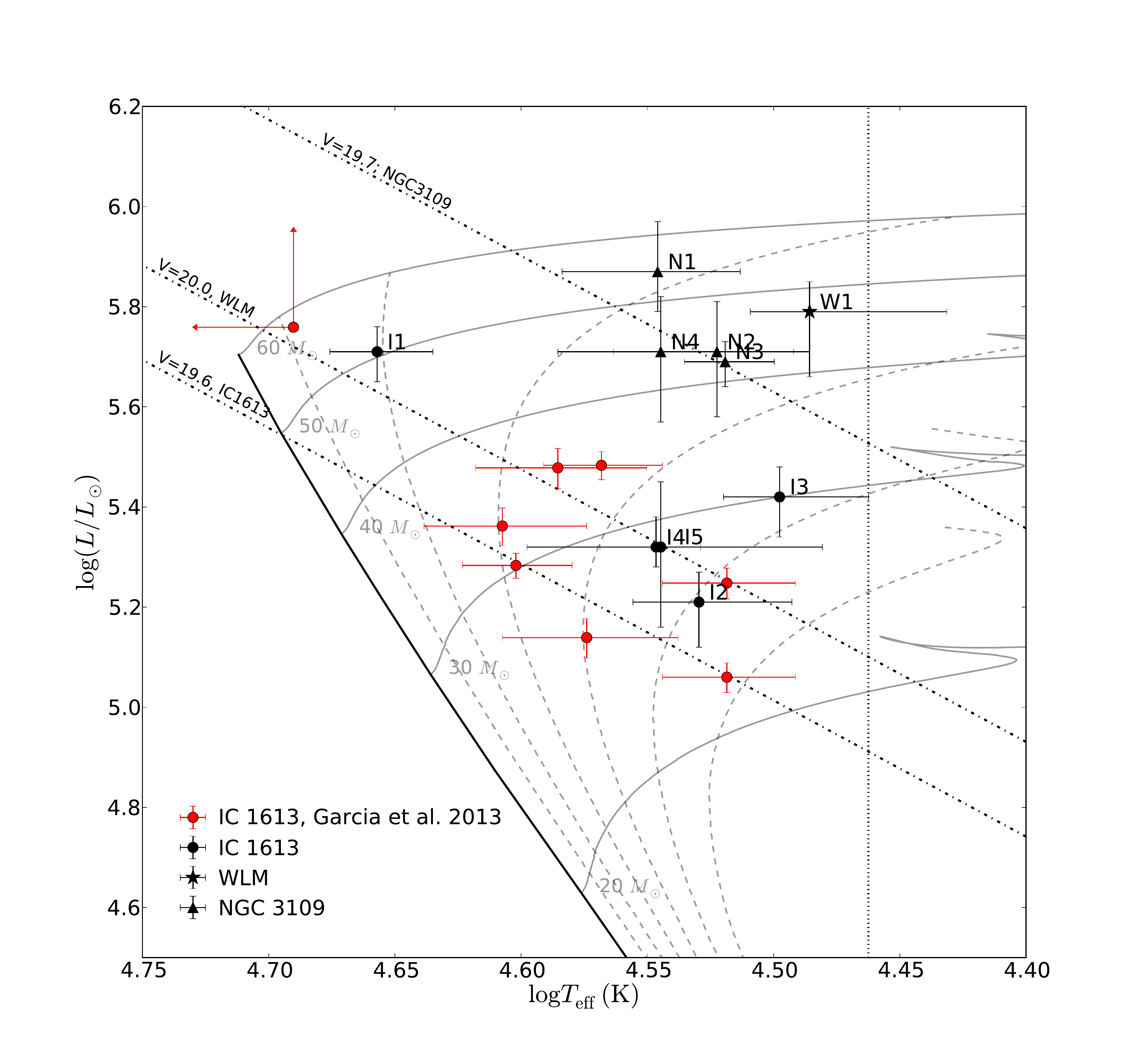}}
  \caption{Herzsprung-Russell diagram indicating the location of the target stars. Also plotted are evolutionary tracks from \cite{brott2011} for SMC metallicity and no initial rotation. The dashed lines indicate isochrones in steps of 1 Myr. The dashed-dotted lines indicate the magnitude cut-offs for the three galaxies, and the dotted vertical line roughly indicates the division between O and B stars. The IC1613 stars from \cite{garcia2013} are plotted in red.}
  \label{fig:hrd}
\end{figure}

\par Figure~\ref{fig:hrd} shows the position of the full sample of stars in the Herzsprung-Russell diagram (HRD). Our sample is complete for the O stars listed in \citet{bresolin2006, bresolin2007},  and \citet{evans2007} above the indicated magnitude cut-off for each galaxy. \cite{garcia2013} identified eight new O-type stars in IC1613, and provided an estimate of their stellar parameters. We also show these stars in Figure~\ref{fig:hrd}, but note that they are based on observations with a lower resolving power ($R=1000$). \cite{garcia2013} do not give bolometric luminosities, and the values used in the HRD are based on their temperatures and the bolometric correction from \cite{martins2005}.
\par The single WLM star in our sample is at the location in the HRD that is expected for its spectral type. It is remarkable that no other O stars are known in WLM that populate the area of the HRD below WLM-1 and above our magnitude cut-off (indicated with the dashed line in Figure~\ref{fig:hrd}). The only other known O star in WLM is an O7 V((f)) with V=20.36 \citep[A15 in][]{bresolin2006}. This suggests that, while star formation is ongoing in WLM, this mostly happens in low-mass clusters that do not produce many O-type stars. 
\par For NGC~3109, our sample is restricted by the magnitude cut-off. The stars in our sample populate the small area of the HRD that we can observe with X-Shooter, and thus all have high masses ($M_{\mathrm{ini}} \ga 40 M_{\odot}$). They are located in different regions within the host galaxy (see Figure~\ref{fig:ngc3109}), suggesting that massive star formation is ongoing in several regions of the galaxy.
\par The HRD for IC~1613 is well populated by our sample and the stars from \cite{garcia2013}. Most of the stars have masses in the range $25 M_{\odot} \la M_{\mathrm{ini}} \la 35 M_{\odot}$, but the two O3 stars indicate that higher mass stars are also being formed. This is further suggested by the presence of the oxygen sequence Wolf-Rayet star in the galaxy \citep[DR1; see, e.g., ][]{kingsburgh1995, tramper2013}. While on large time-scales the star-formation rate has been constant \citep{skillman2014}, IC~1613 is currently rigorously forming stars, with 164 OB associations identified \citep{garcia2009}. The location of our sample of stars in IC~1613 follows the main regions of star formation, with the most massive star located in the North-Eastern lobe where star formation is the most prominent. 

\section{Summary}\label{sec:summary}

We have presented the results of a quantitative spectroscopic analysis of ten O-type stars located in the Local Group dwarf galaxies IC~1613, WLM and NGC~3109. These galaxies have a sub-SMC metallicity based on their oxygen content. 
\par We derived the wind and atmosphere parameters by adjusting {\sc fastwind} models to the observed line profiles. We derived the fundamental stellar properties (including ages and initial masses) from comparison with evolutionary tracks.
\par We used our results to investigate the effective temperature versus spectral type calibration at (sub-)SMC metallicity. We presented both weighted and unweighted fits to the giants and supergiants, and find no significant offset between a calibration based on SMC data and one based on the full sample of stars in IC1613, WLM and NGC~3109 within the limits imposed by our data quality.
\par We discussed the location of the sample stars in the Herzsprung-Russell diagram. None of our stars have initial masses higher than $\simeq 50 M_{\odot}$. 
\par We presented the modified wind momentum versus luminosity diagram. Instead of (sub-)SMC strengths winds, our results indicate stellar winds reminiscent of an LMC metallicity. We discussed the indications that the iron content of the host galaxies may be higher than initially thought, and is possibly SMC-like. While this would lessen the discrepancy with radiation-driven wind theory, the stellar winds of the stars in our sample remain significantly too strong for their metallicity. UV observations of the stars are needed to firmly constrain the wind properties and investigate the effect of wind clumping and the potential presence of an additional wind driving mechanism.

\bibliographystyle{aa}
\bibliography{lowZ_arxiv}

\end{document}